%% file: main.tex
\documentclass[conference,a4paper]{IEEEtran}

\input{defs}

\title{Efficient Decoding of Gabidulin Codes\\ over Galois Rings}

\author{\IEEEauthorblockN{Sven Puchinger$^1$, Julian Renner$^2$, Antonia Wachter-Zeh$^2$, Jens Zumbr\"agel$^3$}%
\IEEEauthorblockA{%
$^1$Department of Applied Mathematics and Computer Science, Technical University of Denmark (DTU), Denmark\\
$^2$Institute for Communications Engineering, Technical University of Munich (TUM), Germany\\
$^3$Faculty of Computer Science and Mathematics, University of Passau, Germany\\
Email: svepu@dtu.dk, julian.renner@tum.de, antonia.wachter-zeh@tum.de, jens.zumbraegel@uni-passau.de%
\thanks{S.~Puchinger has received funding from the European Union’s Horizon 2020 research and innovation programme under the Marie Skłodowska-Curie grant agreement no.~713683.
J.~Renner and A.~Wachter-Zeh were supported by the European Research Council (ERC) under the European Union’s Horizon 2020 research and innovation programme (grant agreement no.~801434).}}}

\IEEEoverridecommandlockouts

\begin{document}

\maketitle

\begin{abstract}
This paper presents the first decoding algorithm for Gabidulin codes over Galois rings with provable quadratic complexity.
The new method consists of two steps: (1) solving a syndrome-based key equation to obtain the annihilator polynomial of the error and therefore the column space of the error, (2) solving a key equation based on the received word in order to reconstruct the error vector.
This two-step approach became necessary since standard solutions as the Euclidean algorithm do not properly work over rings.
\end{abstract}

\section{Introduction}

Network coding over finite \emph{rings} \cite{wilson2010joint,nazer2011compute,feng2013algebraic,tunali2015lattices,feng2014communication,gorla2017algebraic} may result in more efficient physical-layer network coding schemes in comparison to using finite \emph{fields}. Since rank-metric codes can be applied for error correction in network coding (cf.~\cite{silva_rank_metric_approach} for finite fields), Kamche and Mouaha \cite{kamche2019rank} considered rank-metric codes over finite principal ideal rings. The authors, amongst others, defined Gabidulin codes over rings and designed a Welch-Berlekamp-like decoding algorithm similar to the one over finite fields~\cite{Loidreau_AWelchBerlekampLikeAlgorithm_2006}. This decoding algorithm has to solve a linear system of equations and perform a polynomial division, resulting in an asymptotic complexity~$\mathcal{O}(n^3)$ for a Gabidulin code of length~$n$. 

In order to accelerate the decoding process, different coding-theoretic approaches can be thought of: a Berlekamp--Massey (BM) approach, an approach based on the Euclidean algorithm, or row reduction techniques.
The Euclidean algorithm requires divisions of polynomials such that the degree of the remainder is smaller than the one of the inputs; over rings, this degree reduction does not work if the leading monomial is not a unit. When investigating row reduction techniques, we encountered a similar problem: having to divide rows by a non-unit element. 
In \cite{byrne2002hamming}, a BM-like decoding approach for Reed--Solomon and BCH codes over rings was presented.
However, when decoding Gabidulin codes, a BM-like approach would only accelerate the first step of decoding, namely, finding the \emph{annihilator polynomial} of the error, not the second step which is necessary to find the explicit error vector. This is fundamentally different from Reed--Solomon codes where the second step (finding the \emph{error values}) is easy and efficient.
All these observations forced us to establish a different decoding technique.

In this paper, we investigate a new approach to decode Gabidulin codes over Galois rings efficiently.
Namely, we first solve a syndrome-based key equation with a BM-like approach to obtain the error span polynomial and then set up another type of key equation based on the received word (in the literature also called \emph{Gao} key equation \cite{Gao_ANewDecodingAlgorithm_2002}) to explicitly recover the error vector in an efficient way. 
This therefore leads to the first approach that decodes Gabidulin codes over Galois rings with provable quadratic complexity.

\section{Preliminaries}

\subsection{Galois Rings}

For a given prime~$p$ and positive integers~$r$ and~$s$ we denote by
$\GR(p^r, s)$ the \emph{Galois ring} of characteristic~$p^r$ and
degree~$s$.  It can be defined as the quotient ring $\ZZ_{p^r}[x] /
(f)$, where $f \in \ZZ_{p^r}[x]$ is a polynomial such that its
reduction $f \bmod p$ in $\F_p[x]$ is irreducible of degree~$s$.

The theory of Galois rings can be viewed as a close analog of the
theory of finite fields, which is translated to the realm of finite
commutative \emph{local} rings, i.e., rings with a unique maximal
ideal.  Galois rings may in fact be more intrinsically defined as the
unique separable ring extensions of~$\ZZ_{p^r}$, or equivalently, as
the \emph{unramified} local ring extensions of~$\ZZ_{p^r}$, meaning
that the principal ideal $(p)$ remains the maximal ideal in those
extensions (see \cite[Sec.~14]{mcdonald1974finite}).

Most importantly for the present work is the property of a Galois ring
being a Galois extension of~$\ZZ_{p^r}$, with group of ring automorphisms
isomorphic to the Galois group of the corresponding residue fields.
More precisely, let $R := \GR(p^r, s)$ and $S := \GR(p^r, t)$ be
Galois rings with residue fields $k := \F_{p^s}$ and $K := \F_{p^t}$,
respectively, and let $s \mid t$ so that $R \subseteq S$ is a ring
extension.  Then the Galois group $\Gal_R(S)$ of ring automorphisms
of~$S$ fixing~$R$ corresponds, by a lifting construction, to the
Galois group $\Gal_k(K)$ of field automorphisms of~$K$ fixing~$k$;
hence, the group $\Gal_R(S)$ is isomorphic to a cyclic group of
order~$m$, where $m = \frac t s = \dim_k K$ is the extension degree
(see \cite[Sec.~15]{mcdonald1974finite}).

\subsection{Computing the Galois Group}

For an extension $k \subseteq K$ of finite fields where
$q := \vert k \vert$, the field Galois group $\Gal_k(K)$ is generated
by a $q$-th power Frobenius map.  Likewise, the ring Galois group
$\Gal_R(S)$ of Galois rings $R \subseteq S$ is also generated by an
automorphism $\sigma \colon S \to S$ that can be described by a $q$-th
power $\alpha \mapsto \alpha^q$ of some element $\alpha \in S$ with
$S = R[\alpha]$ (although it does not hold that $\sigma(z) = z^q$ for
all $z \in S$).  Such an element~$\alpha$ can be constructed by the
following procedure.  Let $\overline f \in k[x]$ be some irreducible
polynomial of degree~$m$ defining the field extension $k \subseteq K$,
then there holds $x^{q^m} \!- x = \overline f \cdot \overline g$ for
some $\overline g \in k[x]$ coprime to~$\overline f$.  By Hensel
lifting \cite[Sec.~13]{mcdonald1974finite} there are $f, g \in R[x]$
with $\overline f = f \bmod p$ and $\overline g = g \bmod p$ such that
$x^{q^m} \!- x = f \cdot g$ holds over~$R$.  Then letting
$S := R[x] / (f)$ and $\alpha := [x] \in S$ we construct a generator
$\sigma \colon S \to S$, $\alpha \mapsto \alpha^q$ of $\Gal_R(S)$ as
desired.

\subsection{Polynomials and Skew Polynomials}

In the following let~$R$ be a finite local commutative ring with
maximal ideal~$\maxIdeal$, which is nilpotent.  Moreover, let
$k := R / \maxIdeal$ be the residue field and let $\mu \colon R \to k$
be the canonical map, extended to polynomials $R[x] \to k[x]$.  The
following results can be found in~\cite[Sec.~13]{mcdonald1974finite}.

For a polynomial $f = \sum f_i x^i \in R[x]$ we have:
\begin{enumerate}
\item $f$~is a unit $\,\Leftrightarrow\,$
  $f_0 \in R^*$ and all $f_i \in \maxIdeal$, $i > 0$ $\,\Leftrightarrow\,$
  $\mu f \in k^*$,
\item $f$~is no zero divisor $\,\Leftrightarrow\,$ some
  $f_i \in R^*$ $\,\Leftrightarrow\,$ $\mu f \ne 0$.
\end{enumerate} 

In the case of 2) the polynomial~$f$ is called \emph{primitive}.

\begin{lemma} Let $g \in R[x]$ be a primitive polynomial.
  \begin{enumerate}
  \item There exists a unit $u \in R[x]$ such that $u g$ is monic;
    moreover, $\deg u g = \deg \mu g \le \deg g$.
  \item For $f \in R[x]$ there is “division with remainder”, i.e.,
    there are $q, r \in R[x]$ with $f = q g + r$ and
    $\deg r < \deg g$.
  \end{enumerate} \end{lemma}

Now let $\sigma \in {\rm Aut}(R)$ be a ring automorphism.  We define
the \emph{skew polynomial} ring $R[x; \sigma]$ via the rule
$x r = \sigma(r) x$ for all $r \in R$, extended by addition and
multiplication.  Still one may apply the canonical map
$\mu \colon R[x; \sigma] \to k[x; \overline \sigma]$, with
$\overline \sigma \in {\rm Aut}(k)$ induced by~$\sigma$, and
above remarks remain valid.

For a polynomial $f = \sum_{i=0}^n f_i x^i \in R[x; \sigma]$ of
degree~$n$ we denote by ${\rm lt}(f) := x^n$ the \emph{leading term},
${\rm lc}(f) := f_n$ its \emph{leading coefficient} and ${\rm lm}(f)
:= {\rm lc}(f) {\rm lt}(f) = f_n x^n$ the \emph{leading monomial}.

\subsection{Smith Normal Form and Rank Profile of Modules}

Consider again an extension $\Rq = \GR(p^r, s) \subseteq \Rqm = \GR(p^r, sm)$ of Galois rings.
Let $\maxIdeal$ be the maximal ideal of $\Rq$, which has nilpotency index~$r$. For $a \in \Rq \setminus \{ 0 \}$ the \emph{valuation} $v(a)$ is defined as the unique integer~$v$ with $a \in \maxIdeal^v \setminus \maxIdeal^{v+1}$, and we let $v(0) := r$.

For any matrix $\A \in \Rq^{m \times n}$ there are invertible matrices $\S \in \Rq^{m \times m}$ and $\T \in \Rq^{n \times n}$ such that $\D = \S \A \T \in \Rq^{m \times n}$, where~$\D$ is called the Smith normal form of~$\A$ and is a diagonal matrix with diagonal entries $d_1, \dots, d_{\min\{n,m\}}$ satisfying $0 \le v(d_1) \le \ldots \le v(d_{\min\{n,m\}}) \le r$.
We define $\rk (\A) := |\{ i\in\{1,\hdots,\min\{m,n\}\}: d_i \not = 0 \}|$ and $\frk (\A) := |\{ i \in \{1,\hdots,\min\{m,n\}\} :d_i \text{ is a unit} \}|$ as the rank and the free rank of $\A$, respectively. The same properties hold for matrices over $\Rqm$, where $\maxIdeal$ needs to be replaced by the maximal ideal of $\Rqm$ denoted by $\MaxIdeal$. 

Let $\ve{\gamma}=[\gamma_1,\hdots,\gamma_m]$ denote an ordered basis of $\Rqm$ over~$\Rq$. We define $\extsmallfield_{\gamma} \colon \Rqm^{n} \rightarrow \Rq^{m\times n}$, $\a \mapsto \A$, where $a_j = \sum_{i=1}^{m} A_{i,j} \gamma_{i}$,\ $j \in \{1,\hdots,n\}$ and denote by $\rk_{\Rq}(\a) := \rk(\A)$ and $\frk_{\Rq}(\a) := \frk(\A)$ the \emph{rank norm} and \emph{free rank norm} of~$\a$, respectively.

Let $\Mspace$ denote an $\Rq$-submodule of $\Rqm$ and let $d_1,\dots,d_n$ refer to the diagonal elements of a matrix in Smith normal form with row space $\Mspace$.
Then, we call the polynomial
\begin{align*}
\phi^{\Mspace}(x) := \sum_{i=0}^{r-1} \phi_i^{\Mspace} x^i \in \ZZ[x]/(x^r)
\end{align*}
 the \emph{rank profile of $\Mspace$}, where $\phi^{\Mspace}_i := \left|\left\{j : v(d_j)=i\right\}\right|$.
Note the relationship between (free) rank and the rank profiles
\[ \frk_{\Rq} \Mspace = \phi^{\Mspace}_0 = \phi^{\Mspace}(0) , \quad
  \rk_{\Rq} \Mspace = \sum_{i=0}^{r-1} \phi^{\Mspace}_i = \phi^{\Mspace}(1) . \]

\subsection{Gabidulin Codes}

Let $R \subseteq S$ be Galois rings and let $\sigma \in \Gal_R(S)$ be
a generating automorphism.  For a skew polynomial $f = \sum_{i=0}^n f_i
x^i \in \Rqm[x; \sigma]$ and $s \in \Rqm$ we let $f(s) := f_0 s + f_1
\sigma(s) + \ldots + f_n \sigma^n(s)$.  Denote by $\Rqm[x; \sigma]_{< k}$
the $S$-module of all skew polynomials of degree~$< k$.  We define
Gabidulin codes as in~\cite{kamche2019rank}.

\begin{definition}%
  Let $\g=[g_1,\hdots,g_n]\in\Rqm^n$, where the entries are linearly independent over $\Rq$, and let $0 < k \le n$. A \emph{Gabidulin code} of length $n$, dimension $k$ and support $\g$ is defined by
  \begin{equation*}
   \Gabcode{\g}{k} := \{f(\g): f \in \Rqm[x;\sigma]_{<k}\}.
 \end{equation*}
\end{definition}
In~\cite[Prop.~3.23]{kamche2019rank}, it was shown that the Gabidulin code $\Gabcode{\g}{k}$ has a generator matrix
  $\G = [\sigma^i(g_j)]_{0 \le i < k ,\, 1 \le j \le n}$.
Further, the minimum rank distance of $\Gabcode{\g}{k}$ is $d = n-k+1$ and $\Gabcode{\g}{k}$ are MRD codes, see~\cite[Thm.~3.24]{kamche2019rank}.
\begin{theorem}\cite[Thm.~3.25]{kamche2019rank}
  Let $\g=[g_1,\hdots,g_n]\in\Rqm^n$, where the entries of $\g$ are linearly independent over $\Rq$, and let $k$ be an integer such that $0<k\leq n$. Then there exists a vector $\h = [h_1,\dots,h_n]$, where the entries of $\h$ are linearly independent over $\Rq$, such that
 $\H = [\sigma^i(h_j)]_{0 \le i < n - k ,\, 1 \le j \le n}$
  is a parity-check matrix of $\Gabcode{\g}{k}$.
\end{theorem}

\section{A Skew-Polynomial Variant of the Byrne-Fitzpatrick Algorithm}\label{sec:byrne_fitzpatrick}

Let $\Rqm := \GR(p^r, t)$ be a Galois ring and let $\sigma \in
{\rm Aut}(S)$ be an automorphism of~$S$.

In order to solve the decoding problem of rank metric Gabidulin codes
over~$S$, following~\cite{fitzpatrick1995key, byrne2002hamming} we
introduce the \emph{solution module} over the skew polynomial ring
$S[x; \sigma]$.  Given a positive integer~$m$ and a polynomial
$u \in S[x; \sigma]$ we let
\[ \mathcal{M} := \left\{ (f, g) \in S[x; \sigma]^2 \mid f u \equiv
  g \bmod x^m \right\} , \]
which is a left submodule of $S[x; \sigma]^2$ (note that the
congruence $\bmod\, x^m$ does not depend on taking left or right
modulo).

Suitable elements of the solution module of minimal degree may be
found by adapting the Gröbner basis approach of Byrne and
Fitzpatrick~\cite{byrne2002hamming} (see also~\cite{byrne2013algebraic}
for a more elementary description for codes over~$\ZZ_4$), as described
next.

We consider a \emph{term order}~$\prec$ on the set of all terms
$\{ (x^n, 0) \mid n \in \NN \} \cup \{ (0, x^n) \mid n \in \NN \}$ of
$S[x; \sigma]^2$, compatible with multiplication by $x^k \in
S[x; \sigma]$ for $k \in \NN$.  Accordingly, for any nonzero pair
in $S[X; \sigma]^2$ the leading term, leading coefficient and
leading monomial can be defined with respect to~$\prec$.  Concretely,
we are going to use the term order given by $(1, 0) \prec (0, 1)
\prec (x, 0) \prec (0, x) \prec \dots$.

A \emph{left Gröbner basis} of the module~$\mathcal{M}$ is a generating
set $\{ (f_i, g_i) \mid i \in I \}$ of~$\mathcal{M}$ such that for all
$(f, g) \in \mathcal{M}$ there exists some $i \in I$ such that
${\rm lm}(f_i, g_i)$ left-divides ${\rm lm}(f, g)$.  Since $(x^m, 0)$
and $(0, x^m)$ are in the solution module~$\mathcal{M}$, by adapting
an argument in~\cite{byrne2002hamming} one can show that~$\mathcal{M}$
has a left Gröbner basis of the form
\[ \mathcal{B} = \{ (a_0, b_0), \dots, (a_{r-1}, b_{r-1}),
  (c_0, d_0), \dots, (c_{r-1}, d_{r-1}) \} \]
with ${\rm lm}(a_i, b_i) = (p^i x^{\lambda_i}, 0)$ and ${\rm lm}(c_j, d_j)
= (0, p^j x^{\mu_j})$ for all $0 \le i, j < r$, for some decreasing
sequences $\lambda_0 \ge \dots \ge \lambda_{r-1}$ and $\mu_0 \ge
\dots \ge \mu_{r-1}$, called \emph{minimal exponents}.

The following algorithm, derived from the method of “solution by
approximations” of~\cite{byrne2002hamming}, efficiently computes a
left Gröbner basis of the solution module~$\mathcal{M}$.

\begin{algorithm}[ht!]
\caption{$\mathsf{SkewByrneFitzpatrick}$}\label{alg:skewbyrnefitzpatrick}
\SetKwInOut{Input}{Input}
\SetKwInOut{Output}{Output}
\Input{$u \in \SkewPolys$ and $m \in \ZZ_{>0}$}
\Output{Left Gröbner basis of the left $\SkewPolys$ module
\begin{align*}
\mathcal{M} := \left\{ (f, g) \in \SkewPolys^2 \mid f u \equiv g \bmod x^m \right\}.
\end{align*}}
let $\mathcal{B}_0 := \left\{ (p^i, 0) \mid i \in \{ 0, \dots, r \!-\! 1 \} \right\} \cup \left\{ (0, p^i) \mid i \in \{ 0, \dots, r \!-\! 1 \} \right\}$ \\
\For{$k \in \{ 0, \dots, m \!-\! 1 \}$}
{\For{each $(f_i, g_i) \in \mathcal{B}_k$}
  {compute the discrepancy $\zeta_i := (f_i u - g_i)_k$ (where $(\cdot)_k$ denotes the $k$-th coefficient)}
  \For{each $(f_i, g_i) \in \mathcal{B}_k$}
  {\If{$\zeta_i = 0$}{put $(f_i, g_i) \in \mathcal{B}_{k+1}$ \\ continue}
    \eIf{there is $(f_j, g_j) \in \mathcal{B}_k$ with ${\rm lt}(f_j, g_j) \prec {\rm lt}(f_i, g_i)$ and $\zeta_j$ divides $\zeta_i$}
    {put $(f_i, g_i) - q (f_j, g_j)$ in $\mathcal{B}_{k+1}$, where $q \in S$ with $\zeta_i = q \zeta_j$}
    {put $(x f_i, x g_i)$ in $\mathcal{B}_{k+1}$}}}
\Return{$\mathcal{B}_m$}
\end{algorithm}

\begin{theorem}\label{thm:skew_Byrne_Fitzpatrick}
  After completing step~$k$ in Algorithm~\ref{alg:skewbyrnefitzpatrick} the
  set~$\mathcal{B}_{k+1}$ is a left Gröbner basis of the module
  $\mathcal{M}_{k+1} := \left\{ (f, g) \in S[x; \sigma]^2 \mid f u \equiv g \bmod x^{k+1} \right\}$.
  In particular, the algorithm is correct.

  It has complexity $O(r m^2)$ operations in~$S$. Furthermore, we have $|\mathcal{B}| = 2r$.
\end{theorem}

\begin{IEEEproof}
  The correctness is proved by induction on~$k$, by adapting the
  arguments in~\cite{byrne2002hamming}.  We briefly sketch it here.
  Let $(\tilde f_i, \tilde g_i)$ be put in~$\mathcal{B}_{k+1}$ in
  $\ell 7$, $\ell 10$ or $\ell 12$ of the algorithm.  First we claim
  that $(\tilde f_i, \tilde g_i) \in \mathcal{M}_{k+1}$, for which we
  show that $(\tilde f_i, \tilde g_i) \in \mathcal{M}_k$ and the
  discrepancy $(\tilde f_i u - \tilde g_i)_k$ vanishes.

  This is obvious in the case of $\ell 7$.  In $\ell 10$ we have
  $(\tilde f_i, \tilde g_i) \in \mathcal{M}_k$, since $(f_i, g_i) ,\,
  (f_j, g_j) \in \mathcal{M}_k$ and $\mathcal{M}_k$ is an $S$-module;
  moreover we have $(\tilde f_i u - \tilde g_i)_k = (f_i u - g_i)_k
  - (q f_j u - q g_j)_k = (f_i u - g_i)_k - q (f_j u - g_j)_k
  = \zeta_i - q \zeta_j = 0$.  And in $\ell 12$ it is clear that
  $(\tilde f_i, \tilde g_i) \in \mathcal{M}_{k+1}$, since $x^k
  \mid f_i u - g_i$ implies $x^{k+1} \mid x f_i u - x g_i$.
  
  Now let $\lambda_0 \ge \ldots \ge \lambda_{r-1}$ and $\mu_0 \ge \ldots
  \ge \mu_{r-1}$, as well as $\lambda_0' \ge \ldots \ge \lambda_{r-1}'$
  and $\mu_0' \ge \ldots \ge \mu_{r-1}'$ be the minimal exponents of
  $\mathcal{M}_k$ and $\mathcal{M}_{k+1}$, respectively.  From the
  inclusions $x \mathcal{M}_k \subseteq \mathcal{M}_{k+1} \subseteq
  \mathcal{M}_k$ we easily infer that
  \begin{equation}\label{eq:proof1}
    \lambda_i \le \lambda_i' \le \lambda_{i+1} \quad\text{ and }\quad
    \mu_j \le \mu_j' \le \mu_{j+1}
  \end{equation} for all $0 \le i, j < r$.

  Suppose that ${\rm lm} (f_i, g_i) = (p^i x^{\lambda_i}, 0)$, then we
  claim that
  \begin{equation}\label{eq:proof2}
    \text{if-condition of~$\ell 9$ holds} \quad\Longleftrightarrow\quad
    \lambda_i = \lambda_i'
  \end{equation} (and a similar statement holds if ${\rm lm} (f_j, g_j)
  = (0, p^j x^{\mu_j})$).  Indeed, if $\ell 9$ holds, then ${\rm lm}
  (\tilde f_i, \tilde g_i) = {\rm lm} (f_i, g_i)$, so that $\lambda_i
  = \lambda_i'$.  Conversely, suppose that $\lambda_i = \lambda_i'$,
  thus there is $(\tilde a, \tilde b) \in \mathcal{M}_{k+1}$ such
  that ${\rm lm}(\tilde a, \tilde b) = (p^i x^{\lambda_i}, 0)$, and
  hence we have $(\tilde a, \tilde b) - (f_i, g_i) \in \mathcal{M}_k$
  with ${\rm lt}( (\tilde a, \tilde b) - (f_i, g_i) ) \prec
  (x^{\lambda_i}, 0)$.  By the division algorithm we may write
  $(\tilde a, \tilde b) - (f_i, g_i) = \sum \alpha_l (a_l, b_l)
  + \sum \beta_l (c_l, d_l)$ with $\alpha_l, \beta_l \in S[x; \sigma]$
  and ${\rm lt} (a_l, b_l) ,\, {\rm lt} (c_l, d_l) \prec (x^{\lambda_i}, 0)$.
  For the discrepancy we then find $0 = \zeta_i - q \zeta_j$ for some
  $q \in S$ and some~$j$ with leading term less than $(x^{\lambda_i}, 0)$.
  Therefore, the condition in $\ell 9$ is satisfied.

  From~(\ref{eq:proof1}) and~(\ref{eq:proof2}) it follows that if
  $\mathcal{B}_k$ is a Gröbner basis of~$\mathcal{M}_k$, then
  $\mathcal{B}_{k+1}$ as produced by Algorithm~\ref{alg:skewbyrnefitzpatrick}
  is a Gröbner basis of~$\mathcal{M}_{k+1}$, establishing the correctness.

  For the running time analysis, observe first that there are $2 r$
  pairs $(f_i, g_i)$ in the Gröbner bases~$\mathcal{B}_k$, and the
  degree of the polynomials $f_i, g_i$ is in $O(m)$ as it increases in
  each outer loop by at most~$1$.  Hence the computation of each
  discrepancy in~$\ell 4$ requires $O(m)$ operations in~$S$.  The
  if-condition in~$\ell 9$ can easily be checked by considering the
  degrees and valuations; neglecting this cost we only take $\ell 10$
  into account, which again needs $O(m)$ operations in~$S$.
  Therefore, completing one step~$k$ of the outer loop amounts to
  $O(r m)$ operations in~$S$, which results in the stated overall
  unning time.
\end{IEEEproof}

\section{A New Decoder for Gabidulin Codes over Galois Rings}\label{sec:decoder}

In this section, we propose a new decoding algorithm for Gabidulin codes over rings with quadratic complexity in the code length.
The first part of the decoder is to retrieve a skew polynomial called \emph{annihilator polynomial}, which vanishes on the module spanned by the error vector.
In the literature, this polynomial is also called \emph{error span polynomial}.
We obtain this by solving a syndrome-based key equation via the skew Byrne--Fitzpatrick algorithm presented in the previous section.

The second part of the algorithm uses a different kind of key equation, which involves the message polynomial of the transmitted codeword, to retrieve this message polynomial under the condition that the rank of the error is small. This is done using standard operations with skew polynomials, such as interpolation and left and right division.

\begin{definition}
Let $\e \in \Rqm^n$. An \emph{annihilator polynomial} of $\e$ is a primitive polynomial $\Lambda \in \SkewPolys$ of minimal degree
such that $\Lambda(e_i)=0$ for all $i=1,\dots,n$.
\end{definition}

\begin{lemma}\label{lem:MSP_degree_and_number}
Let $\e \in \Rqm^n$. %
Any annihilator polynomial has degree exactly $t := \rank(\e)$. %
Moreover, if $\rk(\e) = \frk(\e)$, then there is a unique monic annihilator polynomial of $\e$.
\end{lemma}

\begin{IEEEproof}
By \cite[Prop.~2.5]{kamche2019rank} there exists a monic (hence primitive) polynomial of degree $t$ that vanishes on the $e_i$. This implies that an annihilator polynomial has degree at most $t$.
Furthermore, by \cite[Prop.~3.16]{kamche2019rank}, any polynomial of degree $<t$ that vanishes on the $e_i$ cannot be primitive, which proves that the degree must be at least $t$.
The second claim directly follows from \cite{kamche2019rank}.
\end{IEEEproof}

We need the following lemma to derive the key equation that we use for decoding.
The statement generalizes the decomposition of the error's matrix representation, which was already used for decoding in \cite{ga85a}. The difference is that, over rings, the entries of $\a$ are not necessarily linearly independent, but the rank profile of $\a$ coincides with the rank profile of $\e$.

\begin{lemma}\label{lem:decomposition}
Let $\e \in \Rqm^n$ and define $t := \rank(\e)$.
Then there is a vector $\a \in \Rqm^t$ with the same rank profile as $\e$ and a matrix $\B \in \Rq^{t \times n}$ whose rows are linearly independent, such that
\begin{align*}
\e = \a \B.
\end{align*}
The entries of $\a$ form a minimal generating set of $\langle e_1,\dots,e_n\rangle$.
\end{lemma}

\begin{IEEEproof}
Expand $\e \in \Rqm^n$ into a matrix $\E \in \Rq^{r \times n}$. By the existence of the Smith normal form, we can decompose
$\E = \A' \D' \B'$,
where $\A' \in \Rq^{r \times r}$ and $\B' \in \Rq^{n \times n}$ are invertible matrices and $\D' \in \Rq^{r \times n}$ is a diagonal matrix with diagonal entries
\begin{equation*}
p^{i_1}, \dots, p^{i_t}, 0, \dots, 0
\end{equation*}
with $\min \{ n, r \} - t$ many zeros,
where the powers $0 \leq i_j<r$ correspond to the rank profile of $\E$ (which is the same as the one of $\e$).
Due to the $\min\{n,r\}-t$ zero entries on the diagonal of $\D$, we can write
$\E = \tilde{\A} \D \B$,
where $\tilde{\A}$ consists of the first $t$ columns of $\A$, $\D$ is the left-upper $t \times t$ submatrix of $\D'$, and $\B$ consists of the first $t$ rows of $\B'$.
Note that the columns of $\tilde{\A}$ and the rows of $\B$ are linearly independent.
Define $\A := \tilde{\A} \D \in \Rq^{r \times n}$ and observe that $\A$ has the same rank profile as $\E$.
We obtain $\a$ as in the claim by writing every column of $\A$ as an element of $\Rqm$.
\end{IEEEproof}

For a received word $\r \in \Rqm^n$, we define the \emph{syndrome polynomial}
\begin{equation}
s_\r(x) := \sum_{i=0}^{\!\!n-k-1\!\!} \Big( \sum_{j=1}^{n} \sigma^i(h_j) r_j \Big) x^i \in \SkewPolys_{< n-k}, \label{eq:syndrome}
\end{equation}
where $h_1,\dots,h_n \in \Rqm$ corresponds to the first row of the parity-check matrix. Note that $s_\c = 0$ for all $\c \in \Code$, so if $\r = \c+\e$, the syndrome polynomial only depends on the error $\s_\r = \s_\e$.

Our decoding algorithm solves the key equation, i.e., it finds polynomials $[\lambda,\omega]$, which fulfill the same congruence relation as $\Lambda$ and $\Omega$ and satisfy the same degree constraints.

\begin{theorem}[Syndrome Key Equation]\label{lem:syndrome_key_equation}
Let $\Lambda$ be an annihilator polynomial of $\e$. Then, there is a skew polynomial $\Omega$ of degree $\deg \Omega < \deg \Lambda$ such that
\begin{align*}
\Lambda s_\e \equiv \Omega \bmod x^{n-k},
\end{align*}
where $s_\e \in \SkewPolys_{< n-k}$ is the syndrome polynomial of $\e$, as defined in \eqref{eq:syndrome}.
\end{theorem}

\begin{IEEEproof}
Recall from Lemma~\ref{lem:decomposition} that $\e = \a \B$ where $\a = [a_1,\dots,a_t]$ and $\B = [B_{l,j}]_{1\leq l \leq t ,\, 1 \leq j \leq n}$.
We define
$d_l :=  \sum_{j=1}^{n}B_{l,j}h_j$.
The coefficients of the syndrome $s_\e = [s_{\e, 1}, \dots, s_{\e, {n-k}}]$ are
\begin{align*}
s_{\e,i} &= \sum_{j=1}^{n} e_j \sigma^i(h_j) = \sum_{j=1}^{n}\sum_{l=1}^{t}a_l B_{l,j}\sigma^i(h_j)
 = \sum_{l=0}^{t-1} a_l \sigma^i(d_l),
\end{align*}
for all $i = 1, \dots, n\!-\!k$. 
The $i$-th coefficient of $\Lambda s_\e$, where $i = 0, \dots, n\!-\!k\!-\!1$, can be calculated by
\begin{align*}
\Omega_i &:= \big[\Lambda s_\e\big]_i = \sum_{j=0}^{i} \Lambda_j \sigma^j(s_{\e,i-j})\\ 
&= \sum_{j=0}^{i} \Lambda_j \sigma^j \Big( \sum_{l=1}^{t} a_l \sigma^{i-j}(d_l) \Big)
= \sum_{l=1}^{t}\sigma^i(d_l)  \sum_{j=0}^{i} \Lambda_j  \sigma^j(a_l).
\end{align*}
For any $i \geq t$ this gives:
\begin{equation*}
\Omega_i = \sum_{l=1}^{t}\sigma^i(d_l) \Lambda\big(a_l\big) = 0 , \quad \text{for all } i = t, \dots, {n\!-\!k\!-\!1},
\end{equation*}
since by definition $\Lambda(x)$ has~$a_i$ as roots, for all $i = 1, \dots, t$, and therefore $\deg \Omega< t = \deg\Lambda$.
\end{IEEEproof}

We will use the following theorem to show how to retrieve (under the condition that the error has small rank) the message polynomial of the transmitted codeword from the output of the skew Byrne--Fitzpatrick algorithm.

\begin{theorem}\label{thm:decoder_correctness_second_part}
Let $\r = \c + \e$, where $\c \in \Code$ with message polynomial $f$ and $t := \rank(\e) \leq \tfrac{n-k}{2}$.
Let $s = s_\r = s_\e$ be the syndrome polynomial corresponding to $\r$. %
Suppose that we have two non-zero polynomials $u,v \in \SkewPolys$ such that:
\begin{itemize}
\item $u$ is primitive
\item $u s - v \equiv 0 \bmod x^{n-k}$
\item $\deg u \leq t$
\item $\deg v < \deg u$
\end{itemize}
Then $u$ is an annihilator polynomial of $\e$.
In particular, its degree equals~$t$.
Furthermore, we have
\begin{equation*}
u R \equiv u f \modr G,
\end{equation*}
where~$R$ is the unique interpolation polynomial of $\r$, and~$G$ is the (unique, since the $g_i$ are linearly independent) annihilator polynomial of the~$g_i$ (which has degree~$n$).
\end{theorem}

\begin{IEEEproof}
Due to $\deg v < t$, $2t \!-\! 1 < n \!-\! k$ and the congruence $u s - v \equiv 0 \bmod x^{n-k}$, the skew polynomial $u$ satisfies
\begin{align*}
(u s)_i = 0, \quad \text{for all } i=t,\dots,2t\!-\!1,
\end{align*}
where $(u s)_i$ denotes the $i$-th coefficient. Written as a linear system in the coefficients $u_0,\dots,u_t$ of $u$, we get
\begin{align*}
\underbrace{\begin{bmatrix}
\sigma^0(s_t) & \sigma^1(s_{t-1}) & \dots & \sigma^t(s_{0}) \\
\sigma^0(s_{t+1}) & \sigma^1(s_{t}) & \dots & \sigma^t(s_{1}) \\
\vdots & \vdots & \ddots & \vdots \\
\sigma^0(s_{2t-1}) & \sigma^1(s_{2t-2}) & \dots & \sigma^t(s_{t-1}) \\
\end{bmatrix}
}_{=: \, \S}
\begin{bmatrix}
u_0 \\
u_1 \\
\vdots \\
u_t
\end{bmatrix}
=
\begin{bmatrix}
0 \\
0 \\
\vdots \\
0
\end{bmatrix}
\end{align*}
Due to Lemma~\ref{lem:decomposition}, there is an $\a \in \Rqm^t$ with the same rank profile as $\e$ and a matrix $\B \in \Rq^{t \times n}$ whose rows are linearly independent, such that
$\e = \a \B$,
and the entries of $\a$ are a minimal generating set of $\langle e_1,\dots,e_n\rangle$.
Define
\begin{align*}
\d = [d_1,\dots,d_t] := \h \B^\top
\end{align*}
and observe that the entries of $\d$ are linearly independent over~$\Rq$, since the both the entries of $\h$ and the rows of $\B$ are linearly independent.
As in \cite{gabidulin1991fast}, we can decompose the matrix $\S$ as follows:
\begin{align*}
  \S = \D \A^\top,
\end{align*}
with $\D:= [\sigma^{t+i}(d_j)]_{0\leq i<t , 1\leq j\leq t}$, $\A:= [\sigma^{i}(a_j)]_{0\leq i\leq t , 1\leq j\leq t}$.
Since the $d_i$ are linearly independent over $\Rq$, so are the $\sigma^t(d_i)$. Hence, the square Moore matrix $\D$ is invertible and $u$ simply satisfies the linear system
\begin{align*}
\A^\top
\begin{bmatrix}
u_0 \\
\vdots \\
u_t
\end{bmatrix}
= \0.
\end{align*}
This can be rewritten as $u(a_i)=0$ for all $i=1,\dots,t$. Since the $a_i$ are a generating set of $\langle e_1,\dots,e_n\rangle$, by the linearity of the skew polynomial evaluation, we get that $u(e_i)=0$ for all $i=1,\dots,n$.
Furthermore, $u$ is primitive and of minimal degree ($=t$) among all monic polynomials with this property due to Lemma~\ref{lem:MSP_degree_and_number}. This proves that $u$ is an annihilator polynomial.

The second part of the claim follows directly since
\begin{align*}
\big[u(R \!-\! f)\big](g_i) &= u\big(R(g_i) \!-\! f(g_i)\big) 
= u\big(r_i \!-\! c_i\big) 
= u\big(e_i\big) 
= 0,
\end{align*}
where the last equality follows from the first part. Since the $g_i$ are linearly independent, we have that $G$ must right-divide $u(R-f)$.
\end{IEEEproof}

We need one last lemma, which shows that all the skew polynomial operations needed for the new decoder can be implemented in quadratic complexity.

\begin{lemma}\label{lem:fast_operations}
Let $f,g \in \SkewPolys_{\leq n}$.
The following operations with skew polynomials over $\Rqm$ can be implemented in $O(n^2)$ operations in $\Rqm$:
\begin{enumerate}
\item \label{itm:fast_multiplication} Multiplication $ab$, where $a,b \in \SkewPolys_{\leq n}$.
\item \label{itm:fast_division} Left and right division of $a$ by $b$, where $a,b \in \SkewPolys_{\leq n}$ and $b$ is primitive.
\item \label{itm:fast_interpolation} Computing the unique interpolation polynomial of $\{(g_i,r_i)\}_{i=1}^{n}$, where the $g_i \in \Rqm$ are linearly independent over $\Rq$ and the $r_i \in \Rqm$ are arbitrary.
\item \label{itm:fast_annihilator} Computing a monic annihilator polynomial of $\g$, where $\g \in \Rqm^n$.
\item \label{itm:fast_MPE} Computing $[a(g_1),\dots,a(g_n)]$, where $a \in \SkewPolys_{\leq n}$ and $g_1,\dots,g_n \in \Rqm$ are linearly independent over $\Rq$.
\end{enumerate}
\end{lemma}

\begin{IEEEproof}
\ref{itm:fast_multiplication}) is obvious by definition and \ref{itm:fast_annihilator}) follows by carefully analyzing the algorithm given in \cite[Prop.~2.5]{kamche2019rank}.

Ad \ref{itm:fast_MPE}): It costs $O(n)$ operations to evaluate one polynomial of degree at most $n$. Hence, evaluating it at $n$ points naively costs $O(n^2)$ operations in $\Rqm$.

Ad \ref{itm:fast_division}): If $g$ is monic, division works as in the case of finite fields (see, e.g., \cite[Alg.~2.1, Alg.~2.2]{wachter2013decoding}), i.e., in quadratic complexity. If $g$ is not monic, then by the inductive procedure of~\cite[Lem.~13.5]{mcdonald1974finite} one may construct in quadratic time a unit polynomial $u \in \SkewPolys$ such that $ug$ is monic; then divide~$f$ by $ug$. The multiplication $ug$ costs at most $O(n^2)$ operations, and we can use the quadratic decoder for division by a monic polynomial.

Ad \ref{itm:fast_interpolation}): Using the recursive strategy in \cite[Lem.~16]{puchinger2018fast} (which applies as well in the case of rings), one can compute an interpolation polynomial at $n$ points by
\begin{itemize}
\item two interpolations at $\approx n/2$ points,
\item computing two annihilator polynomials of vectors of length $\approx n/2$, and
\item multiplication of two skew polynomials of degree $\approx n/2$.
\end{itemize}
Since the latter two kinds of operations have quadratic complexity, the master theorem implies that the overall complexity of interpolation is quadratic in $n$.
\end{IEEEproof}

\begin{algorithm}[ht!]
\caption{$\mathsf{Decoder}$}\label{alg:decoder}
\SetKwInOut{Input}{Input}
\SetKwInOut{Output}{Output}
\Input{$\r \in \Fqm^n$} %
\Output{%
If there is a $\c = \left[f(g_1),\dots,f(g_n)\right] \in \Code$ with $f \in \SkewPolys_{< k}$ and $\dR(\r,\c) \leq \tfrac{n-k}{2}$, then $f$.\\ Otherwise ``decoding failure''.}

$\s \gets \H \r^\top$ \\
$s \gets \sum_{i=0}^{n-k-1} s_i x^i$ \\
$\mathcal{B} \gets \mathsf{SkewByrneFitzpatrick}(s, n\!-\!k)$ \\
$(\lambda,\omega) \gets$ element of $\mathcal{B}$ of minimal degree among all~$(u,v) \in \mathcal{B}$ with $\deg u > \deg v$ and $u$ primitive \\
$R \gets$ unique interpolation polynomial of $\{(g_i,r_i)\}_{i=1}^{n}$ \label{line:R} \\
$G \gets$ unique annihilator polynomial of the $g_1,\dots,g_n$ \label{line:G} \\
$\psi \gets \lambda R \remr G$ \\
$(f,\rho) \gets$ (quotient, rem.) of left division of $\psi$ by $\lambda$ \\
\If{$\rho =0$ and $\dR\!\left( \r, \, \left[ f(g_1),\dots,f(g_n)\right]\right) \leq \tfrac{n-k}{2}$ and $\deg f < k$\label{line:check_if_f_has_correct_form}}{
\Return{$f$}
} \Else {
\Return{``decoding failure''}
}
\end{algorithm}

\begin{theorem}\label{thm:correctness_and_complexity}
Algorithm~\ref{alg:decoder} is correct and has complexity $O(r n^2)$ operations in $\Rqm$.
\end{theorem}

\begin{IEEEproof}
Assume that there is a codeword $\c$ with message polynomial $f$ and rank distance at most $\tfrac{n-k}{2}$ to the received word.
Define $\e := \r-\c$ and $t := \rank(\e) \leq \tfrac{n-k}{2}$.

The skew Byrne--Fitzpatrick algorithm outputs a left Groebner basis of the module
\begin{equation*}
\mathcal{M} := \left\{ (u,v) \mid u s \equiv v \bmod x^{n-k} \right\}.
\end{equation*}

Hence, the output basis must contain a pair $(u, v) \in \mathcal{M}$ with~$u$ primitive, $\deg u>\deg v$, and $\deg u$ minimal among the pairs with these properties. %

By Lemma~\ref{lem:syndrome_key_equation}, there is a pair $(u,v) = (\Lambda,\Omega)$ with $\deg u = t$ that fulfills the properties above. Thus, the~$u$ of minimal degree has degree at most~$t$.

Hence, since also $t\leq \tfrac{n-k}{2}$, by Theorem~\ref{thm:decoder_correctness_second_part} the polynomial~$u$ is a valid annihilator polynomial of the error $\e$. Moreover,
\begin{equation*}
u R \equiv u f \modr G,
\end{equation*}
where $R$ and $G$ are the unique polynomials computed in Lines~\ref{line:R} and \ref{line:G} of Algorithm~\ref{alg:decoder}, and $f$ is the message polynomial corresponding to the codeword $\c$.

Since $\deg uf =  \deg u + \deg f < t+k-1 < n = \deg G$, we obtain $uf$ by right division of $uR$ by $G$. This division is well-defined since $G$ is monic.

Finally, we obtain the message polynomial $f$ by left division of $uf$ by $u$. This is possible since $u$ is primitive.

If there is no codeword with radius $\tfrac{n-k}{2}$ around the received word, Line~\ref{line:check_if_f_has_correct_form} ensures that the output is ``decoding failure''.

The complexity follows by Theorem~\ref{thm:skew_Byrne_Fitzpatrick} and the discussions on the complexity of operations with skew polynomials in Lemma~\ref{lem:fast_operations}.
\end{IEEEproof}
The proof of the first claim in Theorem~\ref{thm:decoder_correctness_second_part} works similar to its finite field analog (see, e.g., \cite{gabidulin1991fast}). A difference is that we need to take care that we use the correct kind of decomposition $\e = \a \B$ of the error. Furthermore, in the case of finite fields, the obtained $u$ is uniquely determined. Here, the polynomials $u$ that satisfy the conditions of Theorem~\ref{thm:decoder_correctness_second_part} are \emph{all} valid annihilator polynomials of $\e$ (cf.~Lemma~\ref{lem:MSP_degree_and_number} for the number of such polynomials).
In our case, it is advantageous to calculate the message polynomial instead of the error values (as done in \cite{gabidulin1991fast})
for complexity reasons: our method uses only operations with quadratic (or faster) algorithms (cf.~Lemma~\ref{lem:fast_operations}).

On the other hand, we did not directly solve the key equation $\Lambda R \equiv \Lambda f \modr G$ since we rely on an adaptation of the Byrne--Fitzpatrick algorithm, which is only able to solve key equations with moduli of the form $x^i$.

Theorem~\ref{thm:correctness_and_complexity} shows that Algorithm~\ref{alg:decoder} is asymptotically faster than Kamche and Mouaha's Welch--Berlekamp-like decoder \cite[Alg.~1]{kamche2019rank}.
The latter algorithm relies on solving a linear system of equations over $\Rqm$, which costs $O(n^\omega)$ operations in $\Rqm$ using Smith normal form (cf.~\cite{storjohann2000algorithms}), where $2 \leq \omega \leq 3$ is the exponent of the used matrix multiplication algorithm (naive: $\omega=3$, Strassen's algorithm: $\omega\approx2.81$, currently best-known: $\omega \approx 2.37$).
Since there is no complexity analysis of the other decoders in \cite{kamche2019rank},
Theorem~\ref{thm:correctness_and_complexity} beats the previous best-known cost bound on the complexity of decoding Gabidulin codes over Galois rings.

\section{Future Work}

Our proposed decoding algorithm has quadratic complexity. However, the cost bounds in Lemma~\ref{lem:fast_operations} can be reduced to sub-quadratic complexity using the results in \cite{caruso2012some,caruso2017fast,puchinger2018fast} and thus, our approach might be improved such that it has sub-quadratic complexity.

It would be interesting to find a variant of the Byrne--Fitzpatrick algorithm that can solve key equations with arbitrary moduli. This would allow us to solve the key equation $\Lambda R \equiv \Lambda f \modr G$ directly instead of the two-step process.

In~\cite{fitzpatrick1995key}, algorithms of the same forms as the extended Euclidean, the Berlekamp--Massey and the Peterson--Gorenstein--Zierler algorithms were proposed for Galois rings. However, only the latter one was generalized to finite rings. An interesting open problem is the generalization of an extended Euclidean like algorithm to finite rings and to propose a sub-quadratic speed-up.

\section*{Acknowledgement}

We would like to thank Johan Rosenkilde for the valuable discussions.

\newpage
\bibliographystyle{IEEEtran}
\bibliography{main}

\end{document}

%% file: defs.tex
\usepackage{pgfplots}
\pgfplotsset{compat=newest}
\usepackage{tikz}
\usetikzlibrary{arrows,matrix,positioning,patterns}
\usepackage[utf8]{inputenc}
\usepackage[innermargin=.53in,outermargin=.53in,top=.69in,bottom=.69in]{geometry}
\usepackage[english]{babel}
\usepackage[T1]{fontenc}
\usepackage{epsfig}
\usepackage{amsmath, amssymb, amsbsy}
\usepackage{mathdots}
\usepackage[normalem]{ulem}
\usepackage{xspace}
\usepackage[noend]{algpseudocode}
\usepackage[linesnumbered,ruled,vlined,titlenumbered]{algorithm2e}
\usepackage{algorithmicx}
\usepackage{color}
\usepackage{cite}
\usepackage{booktabs}
\usepackage{verbatim}
\usepackage[colorlinks=true,citecolor=blue,linkcolor=blue,urlcolor=blue]{hyperref}
\usepackage{url}
\usepackage{lipsum}
\usepackage{enumitem}
\usepackage{colortbl}
\usepackage{theorem}
\usepackage{flushend}
\usepackage[normalem]{ulem}
\makeatletter
\newcommand\footnoteref[1]{\protected@xdef\@thefnmark{\ref{#1}}\@footnotemark}
\newcommand{\mylabel}[2]{#2\def\@currentlabel{#2}\label{#1}}
\makeatother

\newtheorem{theorem}{Theorem}
\newtheorem{lemma}[theorem]{Lemma}

\newtheorem{definition}{Definition}

\newenvironment{mymatrix}{\begin{bmatrix}} {\end{bmatrix} }

\def\ve#1{{\mathchoice{\mbox{\boldmath$\displaystyle #1$}}%
              {\mbox{\boldmath$\textstyle #1$}}%
              {\mbox{\boldmath$\scriptstyle #1$}}%
              {\mbox{\boldmath$\scriptscriptstyle #1$}}}}

\definecolor{lightcolor}{rgb}{0.8,0.8,0.8}
\definecolor{armygreen}{rgb}{0.29, 0.33, 0.13}

\newcommand{\Fqm}{\ensuremath{\mathbb{F}_{q^m}}}

\DeclareMathOperator{\extsmallfield}{ext}
\DeclareMathOperator{\rank}{rk}

\DeclareMathOperator{\GR}{GR}

\DeclareMathOperator{\Gal}{Gal}

\newcommand{\Code}{\mathcal{C}}

\newcommand{\Gabcode}[2]{\ensuremath{Gab_{#2}(#1)}}

\newcommand{\0}{\ve{0}}
\renewcommand{\S}{\ve{S}}
\renewcommand{\H}{\ve{H}}

\renewcommand{\a}{\ve{a}}

\newcommand{\e}{\ve{e}}

\newcommand{\g}{\ve{g}}

\renewcommand{\c}{\ve{c}}
\renewcommand{\e}{\ve{e}}

\newcommand{\G}{\ve{G}}

\newcommand{\E}{\ve{E}}

\newcommand{\A}{\ve{A}}
\newcommand{\B}{\ve{B}}

\newcommand{\h}{\ve{h}}

\newcommand{\D}{\ve{D}}

\newcommand{\dR}{\mathrm{d}_\mathrm{R}}

\newcommand{\F}{\mathbb{F}}

\newcommand{\NN}{\mathbb{N}}
\newcommand{\ZZ}{\mathbb{Z}}
\newcommand{\frk}{\mathrm{frk}}
\newcommand{\rk}{\mathrm{rk}}
\newcommand{\T}{\ve{T}}

\newcommand{\Rq}{{R}}
\newcommand{\Rqm}{{S}}

\renewcommand{\r}{\ve{r}}
\newcommand{\s}{\ve{s}}

\newcommand{\Mspace}{\mathcal{M}}

\newcommand{\maxIdeal}{\mathfrak{m}}
\newcommand{\MaxIdeal}{\mathfrak{M}}

\newcommand{\SkewPolys}{\Rqm[x;\sigma]}

\newcommand{\modr}{\; \mathrm{mod}_{\mathrm{r}} \;}

\newcommand{\remr}{\; \mathrm{rem}_{\mathrm{r}} \;}

\renewcommand{\d}{\ve{d}}

%% file: main.bbl
\begin{thebibliography}{10}
\providecommand{\url}[1]{#1}
\csname url@samestyle\endcsname
\providecommand{\newblock}{\relax}
\providecommand{\bibinfo}[2]{#2}
\providecommand{\BIBentrySTDinterwordspacing}{\spaceskip=0pt\relax}
\providecommand{\BIBentryALTinterwordstretchfactor}{4}
\providecommand{\BIBentryALTinterwordspacing}{\spaceskip=\fontdimen2\font plus
\BIBentryALTinterwordstretchfactor\fontdimen3\font minus
  \fontdimen4\font\relax}
\providecommand{\BIBforeignlanguage}[2]{{%
\expandafter\ifx\csname l@#1\endcsname\relax
\typeout{** WARNING: IEEEtran.bst: No hyphenation pattern has been}%
\typeout{** loaded for the language `#1'. Using the pattern for}%
\typeout{** the default language instead.}%
\else
\language=\csname l@#1\endcsname
\fi
#2}}
\providecommand{\BIBdecl}{\relax}
\BIBdecl

\bibitem{wilson2010joint}
M.~P. Wilson, K.~Narayanan, H.~D. Pfister, and A.~Sprintson, ``Joint physical
  layer coding and network coding for bidirectional relaying,'' \emph{IEEE
  Transactions on Information Theory}, vol.~56, no.~11, pp. 5641--5654, 2010.

\bibitem{nazer2011compute}
B.~Nazer and M.~Gastpar, ``Compute-and-forward: Harnessing interference through
  structured codes,'' \emph{IEEE Transactions on Information Theory}, vol.~57,
  no.~10, pp. 6463--6486, 2011.

\bibitem{feng2013algebraic}
C.~Feng, D.~Silva, and F.~R. Kschischang, ``An algebraic approach to
  physical-layer network coding,'' \emph{IEEE Transactions on Information
  Theory}, vol.~59, no.~11, pp. 7576--7596, 2013.

\bibitem{tunali2015lattices}
N.~E. Tunali, Y.-C. Huang, J.~J. Boutros, and K.~R. Narayanan, ``Lattices over
  {Eisenstein} integers for compute-and-forward,'' \emph{IEEE Transactions on
  Information Theory}, vol.~61, no.~10, pp. 5306--5321, 2015.

\bibitem{feng2014communication}
C.~Feng, R.~W. N{\'o}brega, F.~R. Kschischang, and D.~Silva, ``Communication
  over finite-chain-ring matrix channels,'' \emph{IEEE Transactions on
  Information Theory}, vol.~60, no.~10, pp. 5899--5917, 2014.

\bibitem{gorla2017algebraic}
E.~Gorla and A.~Ravagnani, ``An algebraic framework for end-to-end
  physical-layer network coding,'' \emph{IEEE Transactions on Information
  Theory}, vol.~64, no.~6, pp. 4480--4495, 2017.

\bibitem{silva_rank_metric_approach}
D.~Silva, F.~R. Kschischang, and R.~K{\"o}tter, ``A rank-metric approach to
  error control in random netw. coding,'' \emph{IEEE Transactions on
  Information Theory}, vol.~54, no.~9, pp. 3951--3967, 2008.

\bibitem{kamche2019rank}
H.~T. Kamche and C.~Mouaha, ``Rank-metric codes over finite principal ideal
  rings and applications,'' \emph{IEEE Transactions on Information Theory},
  vol.~65, no.~12, pp. 7718--7735, 2019.

\bibitem{Loidreau_AWelchBerlekampLikeAlgorithm_2006}
P.~Loidreau, ``{A Welch--Berlekamp Like Algorithm for Decoding Gabidulin
  Codes},'' \emph{Coding and Cryptography --- Revised selected papers of WCC
  2005}, vol. 3969, pp. 36--45, 2006.

\bibitem{byrne2002hamming}
E.~Byrne and P.~Fitzpatrick, ``Hamming metric decoding of alternant codes over
  galois rings,'' \emph{IEEE Transactions on Information Theory}, vol.~48,
  no.~3, pp. 683--694, 2002.

\bibitem{Gao_ANewDecodingAlgorithm_2002}
S.~Gao, ``{A New Algorithm for Decoding Reed--Solomon Codes},'' \emph{Commun.
  Inform. Network Sec.}, vol. 712, pp. 55--68, 2003.

\bibitem{mcdonald1974finite}
B.~R. McDonald, \emph{Finite rings with identity}.\hskip 1em plus 0.5em minus
  0.4em\relax Marcel Dekker Incorporated, 1974, vol.~28.

\bibitem{fitzpatrick1995key}
P.~Fitzpatrick, ``On the key equation,'' \emph{IEEE Transactions on Information
  Theory}, vol.~41, no.~5, pp. 1290--1302, 1995.

\bibitem{byrne2013algebraic}
E.~Byrne, M.~Greferath, J.~Pernas, and J.~Zumbr{\"a}gel, ``Algebraic decoding
  of negacyclic codes over $\mathbb z_4$,'' \emph{Designs, codes and
  cryptography}, vol.~66, no. 1-3, pp. 3--16, 2013.

\bibitem{ga85a}
E.~M. Gabidulin, ``Theory of codes with maximum rank distance,'' \emph{Problemy
  Peredachi Informatsii}, vol.~21, no.~1, pp. 3--16, 1985.

\bibitem{gabidulin1991fast}
------, ``A fast matrix decoding algorithm for rank-error-correcting codes,''
  in \emph{Workshop on Algebraic Coding}.\hskip 1em plus 0.5em minus
  0.4em\relax Springer, 1991, pp. 126--133.

\bibitem{wachter2013decoding}
A.~Wachter-Zeh, ``Decoding of block and convolutional codes in rank metric,''
  Ph.D. dissertation, University of Rennes 1 and Ulm University, 2013.

\bibitem{puchinger2018fast}
S.~Puchinger and A.~{Wachter-Zeh}, ``Fast operations on linearized polynomials
  and their applications in coding theory,'' \emph{Journal of Symbolic
  Computation}, vol.~89, pp. 194--215, 2018.

\bibitem{storjohann2000algorithms}
A.~Storjohann, ``{Algorithms for Matrix Canonical Forms},'' Ph.D. dissertation,
  ETH Zurich, 2000.

\bibitem{caruso2012some}
X.~Caruso and J.~L. Borgne, ``Some algorithms for skew polynomials over finite
  fields,'' \emph{arXiv preprint arXiv:1212.3582}, 2012.

\bibitem{caruso2017fast}
X.~Caruso and J.~Le~Borgne, ``Fast multiplication for skew polynomials,'' in
  \emph{Proceedings of the 2017 ACM on International Symposium on Symbolic and
  Algebraic Computation}, 2017, pp. 77--84.

\end{thebibliography}
